\documentclass[conference,hidelinks]{IEEEtran}

\usepackage{cite}
\usepackage{amsmath,amssymb,amsfonts}
\usepackage{mathtools}
\usepackage{algorithmic}
\usepackage{graphicx}
\usepackage{textcomp}
\usepackage{xcolor}
\def\BibTeX{{\rm B\kern-.05em{\sc i\kern-.025em b}\kern-.08em
    T\kern-.1667em\lower.7ex\hbox{E}\kern-.125emX}}
\usepackage{orcidlink}
\usepackage{glossaries}
\usepackage[outputdir=build,frozencache=true]{minted}
\usepackage{listings}
\usepackage[position=b]{subcaption}
\usepackage{tikz}
\usetikzlibrary{tikzmark,calc,positioning}
\usepackage{enumitem}
\usepackage{hyperref}
\usepackage{cleveref}
\usepackage{makecell}
\usepackage{booktabs}
\usepackage{multirow}
\usepackage[framemethod=TikZ]{mdframed}

\usepackage[T1]{fontenc}
\usepackage{lmodern}

\newacronym{ti}{TI}{thread imbalance}

\newacronym[plural=WANs, firstplural={Wide Area Networks (WANs)}]{wan}{WAN}{Wide Area Network}
\newacronym[plural=WSNs, firstplural={Wireless Sensor Networks (WSNs)}]{wsn}{WSN}{Wireless Sensor Network}
\newacronym{simd}{SIMD}{Single Instruction Multiple Data}
\newacronym{os}{OS}{Operating System}
\newacronym{ble}{BLE}{Bluetooth Low-Energy}
\newacronym{wifi}{Wi-FI}{Wireless Fidelity}
\newacronym[plural=DVS, firstplural={Dynamic Vision Sensors (DVS)}]{dvs}{DVS}{Dynamic Vision Sensor}
\newacronym{ptz}{PTZ}{Pan-Tilt Unit}

\newacronym[plural=FLLs,firstplural=Frequency Locked Loops (FLLs)]{fll}{FLL}{Frequency Locked Loop}
\newacronym{dram}{DRAM}{Dynamic Random Access Memory}
\newacronym{fpu}{FPU}{Floating Point Unit}
\newacronym{fpss}{FPSS}{Floating Point Subsystem}
\newacronym{frep}{FREP}{Floating Point Repetition}
\newacronym{dma}{DMA}{Direct Memory Access}
\newacronym{ssr}{SSR}{Stream Semantic Register}
\newacronym{issr}{ISSR}{Indirection Stream Semantic Register}
\newacronym[plural=LUTs, firstplural={Lookup Tables (LUTs)}]{lut}{LUT}{Lookup Table}
\newacronym[plural=FPGAs, firstplural={Field Programmable Gate Arrays (FPGAs)}]{fpga}{FPGA}{Field Programmable Gate Array}
\newacronym{dsp}{DSP}{Digital Signal Processing}
\newacronym{mcu}{MCU}{Microcontroller Unit}
\newacronym{spi}{SPI}{Serial Peripheral Interface}
\newacronym{cpi}{CPI}{Camera Parallel Interface}
\newacronym{rf}{RF}{Register File}
\newacronym{fifo}{FIFO}{First-In First-Out Queue}
\newacronym{uart}{UART}{Universal Asynchronous Receiver-Transmitter}
\newacronym{raw}{RAW}{Read After Write}
\newacronym[plural=ISAs, firstplural={Instruction Set Architectures (ISAs)}]{isa}{ISA}{Instruction Set Architecture}
\newacronym{xbar}{XBAR}{crossbar}
\newacronym[firstplural=Scratch-Pad Memories (SPMs)]{spm}{SPM}{Scratch-Pad Memory}
\newacronym{ppa}{PPA}{Power Performance Area}
\newacronym{ipi}{IPI}{Inter-Processor Interrupt}
\newacronym[firstplural=Software-Generated Interrupts (SGIs)]{sgi}{SGI}{Software-Generated Interrupt}
\newacronym{pe}{PE}{Processing Element}
\newacronym{tcdm}{TCDM}{Tightly-Coupled Data Memory}
\newacronym{lsu}{LSU}{Load-Store Unit}
\newacronym{icache}{I\$}{Instruction Cache}
\newacronym{dcache}{D\$}{Data Cache}
\newacronym{wfi}{WFI}{Wait For Interrupt}
\newacronym{gpc}{GPC}{GPU Processing Cluster}
\newacronym{cpu}{CPU}{Central Processing Unit}
\newacronym{gpu}{GPU}{Graphics Processing Unit}
\newacronym{llc}{LLC}{Last-Level Cache}
\newacronym{sm}{SM}{Streaming Multiprocessor}
\newacronym[firstplural=Networks on Chip (NoCs)]{noc}{NoC}{Network on Chip}

\newacronym{dfg}{DFG}{Data Flow Graph}
\newacronym{lcg}{LCG}{Linear Congruential Generator}
\newacronym{prn}{PRN}{Pseudo-Random Number}

\newacronym{ste}{STE}{Straight-Through-Estimator}

\newacronym[plural=PTUs, firstplural={Pan-Tilt Units}]{ptu}{PTU}{Pan-Tilt Unit}
\newacronym{mdf}{MDF}{Medium-density fibreboard}
\newacronym{cvat}{CVAT}{Computer Vision Annotation Tool}
\newacronym{coco}{COCO}{Common Objects in Context}
\newacronym{soa}{SoA}{State of the Art}
\newacronym{sf}{SF}{Sensor Fusion}

\newacronym{dl}{DL}{Deep Learning}
\newacronym{bn}{BN}{Batch Normalization}
\newacronym{FGSM}{FBK}{Fast Gradient Sign Method}
\newacronym{lr}{LR}{Learning Rate}
\newacronym{sgd}{SGD}{Stochastic Gradient Descent}
\newacronym{gd}{GD}{Gradient Descent}
\newacronym{llm}{LLM}{Large Language Model}

\newacronym{sta}{STA}{Static Timing Analysis}

\newacronym[plural=GPIOs, firstplural={General Purpose Inupt Outputs (GPIOs)}]{gpio}{GPIO}{General Purpose Input Output}
\newacronym[plural=LDOs, firstplural={Low Dropout Regulators (LDOs)}]{ldo}{LDO}{Low Dropout Regulator}

\newacronym{inq}{INQ}{Incremental Network Quantization}

\newacronym{CV}{CV}{Computer Vision}
\newacronym{EoT}{EoT}{Expectation over Transformation}
\newacronym{RPN}{RPN}{Region Proposal Network}
\newacronym{TV}{TV}{Total Variation}
\newacronym{NPS}{NPS}{Non-Printability Score}
\newacronym{STN}{STN}{Spatial Transformer Network}
\newacronym{MTCNN}{MTCNN}{Multi-Task Convolutional Neural Network}
\newacronym{YOLO}{YOLO}{You Only Look Once}
\newacronym{SSD}{SSD}{Single Shot Detector}
\newacronym{SOTA}{SOTA}{State of the Art}
\newacronym{NMS}{NMS}{Non-Maximum Suppression}
\newacronym{ic}{IC}{Integrated Circuit}
\newacronym{tcxo}{TCXO}{Temperature Controlled Crystal Oscillator}
\newacronym{jtag}{JTAG}{Joint Test Action Group industry standard}
\newacronym{swd}{SWD}{Serial Wire Debug}
\newacronym{sdio}{SDIO}{Serial Data Input Output}

\newacronym[plural=PCBs, firstplural={Printed Circuit Boards (PCB)}]{pcb}{PCB}{Printed Circuit Board}
\newacronym[plural=ASICs, firstplural={Application Specific Integrated Circuits}]{asic}{ASIC}{Application Specific Integrated Circuit}

\newacronym[plural=BNNs, firstplural={Binary Neural Networks (BNNs)}]{bnn}{BNN}{Binary Neural Network}
\newacronym[plural=NNs, firstplural={Neural Networks}]{nn}{NN}{Neural Network (NNs)}
\newacronym[plural=SCMs, firstplural={Standard Cell Memories (SCMs)}]{scm}{SCM}{Standard Cell Memory}
\newacronym{ann}{ANN}{Artificial Neural Networks}
\newacronym{ml}{ML}{Machine Learning}
\newacronym{ai}{AI}{Artificial Intelligence}
\newacronym{iot}{IoT}{Internet of Things}
\newacronym{fft}{FFT}{Fast Fourier Transform}
\newacronym[plural=OCUs, firstplural={Output Channel Compute Units (OCUs)}]{ocu}{OCU}{Output Channel Compute Unit}
\newacronym{alu}{ALU}{Arithmetic Logic Unit}
\newacronym{mac}{MAC}{Multiply-Accumulate}
\newacronym[firstplural={systems-on-chip (SoCs)}]{soc}{SoC}{system-on-chip}
\newacronym[firstplural={multi-processor systems-on-chip (MPSoCs)}]{mpsoc}{MPSoC}{multi-processor system-on-chip}

\newacronym{PGD}{PGD}{Projected Gradient Descend}
\newacronym{CW}{CW}{Carlini-Wagner}
\newacronym{OD}{OD}{Object Detection}

\newacronym{rrf}{RRF}{RADAR Repetition Frequency}
\newacronym{nlp}{NLP}{Natural Language Processing}
\newacronym{qam}{QAM}{Quadrature Amplitude Modulation}
\newacronym{rri}{RRI}{RADAR Repetition Interval}
\newacronym{radar}{RADAR}{Radio Detection and Ranging}
\newacronym{loocv}{LOOCV}{Leave-one-out cross validation}

\newacronym{bsp}{BSP}{Board Support Package}
\newacronym{ttn}{TTN}{The Things Network}
\newacronym{wip}{WIP}{Work in Progress}
\newacronym{json}{JSON}{JavaScript Object Notation}
\newacronym{qat}{QAT}{Quantization-Aware Training}

\newacronym{cls}{CLS}{Classification Error}
\newacronym{loc}{LOC}{Localization Error}
\newacronym{bkgd}{BKGD}{Background Error}
\newacronym{roc}{ROC}{Receiver Operating Characteristic}
\newacronym{frr}{FRR}{False Rejection Rate}
\newacronym{eer}{EER}{Equal Error Rate}
\newacronym{snr}{SNR}{Signal-to-Noise Ratio}
\newacronym{flop}{FLOP}{Floating-Point Operation}
\newacronym{fp}{FP}{Floating-Point}
\newacronym{fps}{FPS}{Frames Per Second}
\newacronym{oi}{OI}{Operational Intensity}
\newacronym{ipc}{IPC}{Instructions per Cycle}

\newacronym{gsc}{GSC}{Google Speech Commands}
\newacronym{mswc}{MSWC}{Multilingual Spoken Words Corpus}
\newacronym{demand}{DEMAND}{Diverse Environments Multichannel Acoustic Noise Database}

\newacronym[plural=SNNs, firstplural={Spiking Neural Networks (SNNs)}]{snn}{SNN}{Spiking Neural Network}
\newacronym[plural=DNNs, firstplural={Deep Neural Networks (DNNs)}]{dnn}{DNN}{Deep Neural Network}
\newacronym[plural=TCNs,firstplural=Temporal Convolutional Networks]{tcn}{TCN}{Temporal Convolutional Network}
\newacronym[plural=CNNs,firstplural=Convolutional Neural Networks (CNNs)]{cnn}{CNN}{Convolutional Neural Network}
\newacronym[plural=TNNs,firstplural=Ternarized Neural Networks]{tnn}{TNN}{Ternarized Neural Network}
\newacronym{ds-cnn}{DS-CNN}{Depthwise Separable Convolutional Neural Network}
\newacronym{rnn}{RNN}{Recurrent Neural Network}
\newacronym{gcn}{GCN}{Graph Convolutional Network}
\newacronym{mhsa}{MHSA}{Multi-Head Self Attention}
\newacronym{crnn}{CRNN}{Convolutional Recurrent Neural Network}
\newacronym{clca}{CLCA}{Convolutional Linear Cross-Attention}

\newacronym{bf}{BF}{Beamforming}
\newacronym{anc}{ANC}{Active Noise Cancellation}
\newacronym{agc}{AGC}{Automatic Gain Control}
\newacronym{se}{SE}{Speech Enhancement}
\newacronym{mct}{MCT}{Multi-Condition Training}
\newacronym{mcta}{MCTA}{Multi-Condition Training \& Adaptation}
\newacronym{pcen}{PCEN}{Per-Channel Energy Normalization}
\newacronym{mfcc}{MFCC}{Mel-Frequency Cepstral Coefficient}
\newacronym{asr}{ASR}{Automated Speech Recognition}
\newacronym{kws}{KWS}{Keyword Spotting}
\newacronym{odl}{ODL}{On-Device Learning}

\newacronym{nl-kws}{NL-KWS}{Noiseless Keyword Spotting}
\newacronym{na-kws}{NA-KWS}{Noise-Aware Keyword Spotting}
\newacronym{odda}{ODDA}{On-Device Domain Adaptation}
\newacronym{hpm}{HPM}{High-Performance Mode}
\newacronym{lpm}{LPM}{Low-Power Mode}

\newcommand{\ResultGeomeanIPCIncrease}{1.62}
\newcommand{\ResultPeakIPC}{1.75}

\newcommand{\ResultPolyLCGIPCIncrease}{1.97}
\newcommand{\ResultMaxPowerIncrease}{1.17}
\newcommand{\ResultGeomeanPowerIncrease}{1.07}
\newcommand{\ResultPeakSpeedup}{2.05}
\newcommand{\ResultGeomeanSpeedup}{1.47}
\newcommand{\ResultPeakEnergySaving}{1.93}
\newcommand{\ResultGeomeanEnergySaving}{1.37}

\setminted[]{
    escapeinside=||,
    style=arduino,
    breaklines
}

\AtBeginDocument{\DeclareCaptionSubType{lstlisting}}

\crefname{step}{step}{steps}
\Crefname{step}{Step}{Steps}
\crefname{type}{type}{types}
\Crefname{type}{Type}{Types}

\glsunset{ipc}

\begin{document}

\newcommand{\edge}[2]{\ensuremath{#1\!\rightarrow\!#2}}

\newcommand\blackcircle[1]{\Circled[inner color=white, outer color=white, fill color=black]{#1}}

\title{Dual-Issue Execution of Mixed Integer and Floating-Point Workloads on Energy-Efficient In-Order RISC-V Cores}

\ifdefined\blindreview
\else
\author{
    \IEEEauthorblockN{Luca Colagrande}
    \IEEEauthorblockA{
        \textit{Integrated Systems Laboratory (IIS)} \\
        \textit{ETH Zurich}\\
        Zurich, Switzerland \\
        colluca@iis.ee.ethz.ch\orcidlink{0000-0002-7986-1975}
    }
    \and
    \IEEEauthorblockN{Luca Benini}
    \IEEEauthorblockA{
        \textit{Integrated Systems Laboratory (IIS)} \\
        \textit{ETH Zurich}\\
        Zurich, Switzerland \\
        lbenini@iis.ee.ethz.ch\orcidlink{0000-0001-8068-3806}
    }
}
\fi

\maketitle

\begin{abstract}
To meet the computational requirements of modern workloads under tight energy constraints, general-purpose accelerator architectures have to integrate an ever-increasing number of extremely area- and energy-efficient processing elements (PEs).
In this context, single-issue in-order cores are commonplace, but lean dual-issue cores could boost PE IPC, especially for the common case of mixed integer and floating-point workloads.
We develop the COPIFT methodology and RISC-V ISA extensions to enable low-cost and flexible dual-issue execution of mixed integer and floating-point instruction sequences. On such kernels, our methodology achieves speedups of 1.47x, reaching a peak 1.75 instructions per cycle, and 1.37x energy improvements on average, over optimized RV32G baselines.
\end{abstract}

\begin{IEEEkeywords}
    RISC-V, dual-issue, energy efficiency, general purpose
\end{IEEEkeywords}

\section{Introduction}

With the end of Dennard scaling, energy-efficiency has become a primary constraint in the design of high-performance computer architectures \cite{horowitz2014}.
Simultaneously, the rapid growth of \gls{ml} models, requiring immense computational resources for training and inference, are dramatically escalating the demand for high-performance computing power \cite{sevilla2022}.

To meet this challenge, modern high-performance architectures integrate an ever-increasing number of cores, designed to meet extreme area- and energy-efficiency targets to enable large-scale replication \cite{choquette2023}.
In-order single-issue cores with shallow pipelines are favoured over power- and resource-hungry deeply-pipelined out-of-order processors.

Within the context of massively-parallel general-purpose accelerators, few architectures propose cores with limited multiple-issue capabilities.
Nvidia's Turing architecture, for instance, implements concurrent execution of FP32 and INT32 operations \cite{burgess2020}, allowing the \gls{fp} math datapath to remain active during the execution of non-\gls{fp}-math instructions.
According to the authors, this results in a 36\% average increase in throughput across several gaming workloads.
Unfortunately, due to the proprietary nature of the architecture, implementation details and their area and power impact remain undisclosed.

A similar form of dual-issue execution was implemented in the Snitch processor \cite{zaruba2021}, coupling a tiny streamlined single-issue in-order RV32I control core and a 64-bit SIMD-capable \gls{fpu}.
Instructions are fetched and issued by the control core, and offloaded to the \gls{fpu} as required.
To support dual-issue execution, the \gls{fpss} integrates a small instruction buffer to cache small \gls{fp} loop bodies, the \gls{frep} buffer.
\gls{fp} instructions are first issued by the integer core, and cached in the \gls{frep} buffer.
In the remaining loop iterations, the \gls{frep} sequencer can independently issue instructions, while the integer core is free to fetch, and potentially execute, successive instructions. Its authors refer to this scheme as \textit{pseudo dual-issue execution}.

This approach implies the following constraints:
\begin{itemize}
    \item the \gls{fp} loop body must be small enough to fit entirely in the \gls{frep} buffer;
    \item the programmer or compiler need to explicitly mark the \gls{fp} loop for \gls{frep} execution in software;
    \item there can be no dependencies between the \gls{fp} and integer threads.
\end{itemize}
The latter constraint follows from the asynchronous execution of the two threads, between which no ordering guarantees are enforced.
While this enables Snitch's dual-issue capabilities at a very low power and area cost, it limits flexibility.
In particular, it hinders the execution of computations featuring a mix of interdependent integer and FP instructions.
We refer to these as mixed integer and \gls{fp} instruction sequences.

In this paper, we present \textit{COPIFT}, a generic method to develop \textbf{C}o-\textbf{O}perative \textbf{P}arallel \textbf{I}nteger and \textbf{F}loating-point \textbf{T}hreads in Snitch, overcoming the limitation imposed by pseudo dual-issue, and enabling low-cost dual-issue execution of mixed integer and \gls{fp} instruction sequences.

To summarize our contributions, we:
\begin{itemize}
    \item Present COPIFT, a generic methodology to enable sustained dual-issue execution of mixed integer and \gls{fp} instruction sequences in Snitch, a state-of-the-art open-source RISC-V processor.
    \item Develop a set of \gls{isa} extensions to broaden the range of applications suited for pseudo dual-issue execution, in conjunction with the COPIFT method.
    \item Implement COPIFT-accelerated codes for selected mixed integer and \gls{fp} instruction sequences, achieving average speedups of \ResultGeomeanSpeedup$\times$ over optimized RV32G baselines, and a peak \gls{ipc} of \ResultPeakIPC.
    \item Present a first account of the energy-efficiency of pseudo multiple-issue execution schemes, demonstrating average energy reductions of \ResultGeomeanEnergySaving$\times$ with COPIFT, over optimized RV32G baselines.
\end{itemize}
Our implementation is fully open source and performance experiments are reproducible using free software.
\ifdefined\blindreview
    \footnote{https://hidden-for-double-blind-review.com}
\else
    \footnote{\url{https://github.com/colluca/snitch\_cluster/tree/copift}}
\fi


\section{Implementation}
\label{sec:implementation}

\subsection{COPIFT Methodology}
\label{sec:copift}

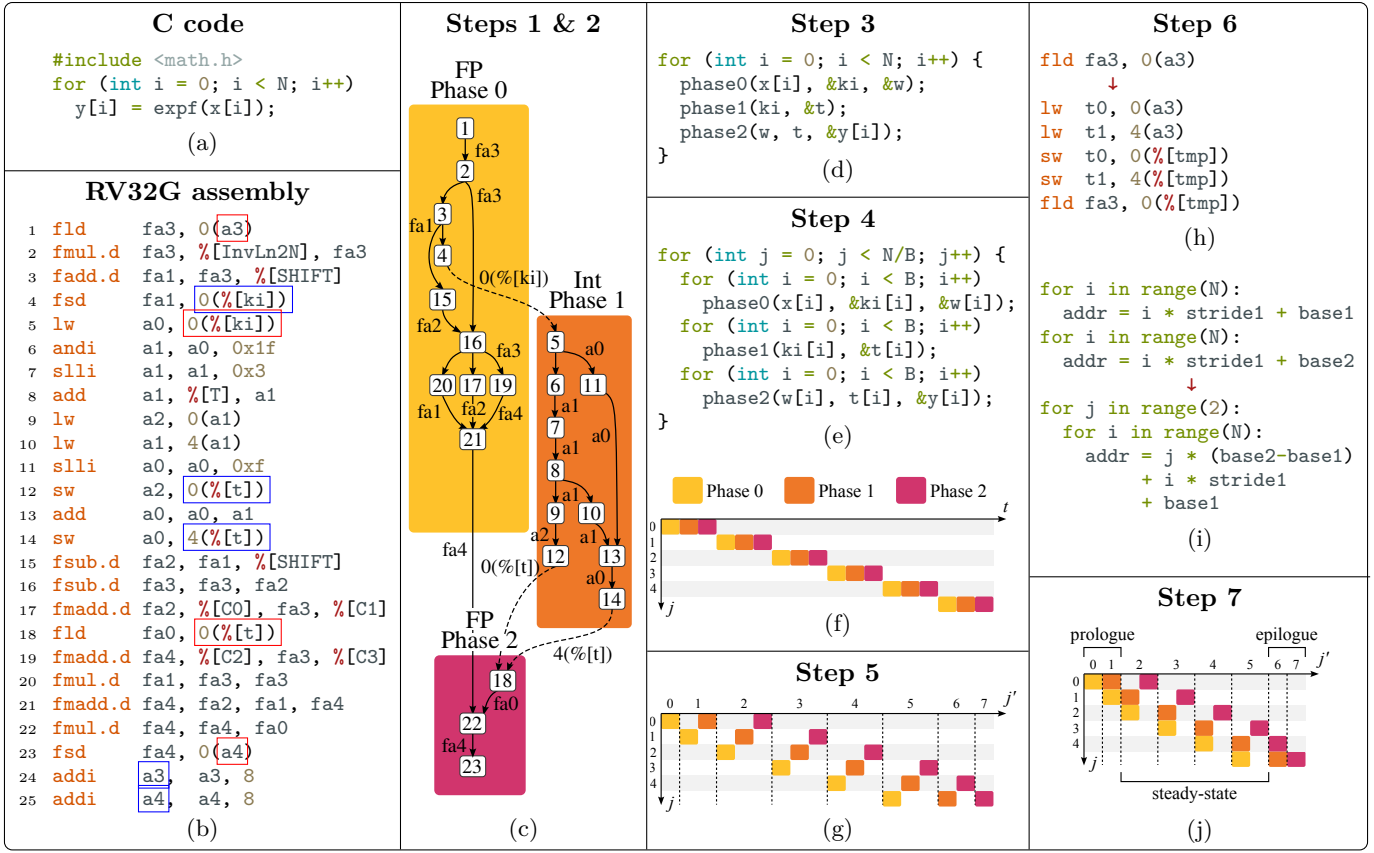
\begin{figure*}[t]
    \begin{mdframed}[innertopmargin=0pt,innerbottommargin=0pt,innerleftmargin=0pt,innerrightmargin=0pt,roundcorner=3pt]
        \begin{minipage}[t]{.29\textwidth}
            \vspace{0pt}
            \begin{minipage}[t]{\textwidth}
                \centering
                \vspace{4pt}
                \textbf{C code}
                \vspace{2pt}
                \begin{minted}[fontsize=\footnotesize,xleftmargin=18pt]{c}
#include <math.h>
for (int i = 0; i < N; i++)
  y[i] = expf(x[i]);
                \end{minted}
                \vspace{-4pt}
                \subcaption{}
                \label{lst:c}
                \vspace{-4pt}
                \hrulefill
            \end{minipage}
            \vfill
            \begin{minipage}[t]{\textwidth}
                \centering
                \vspace{4pt}
                \textbf{RV32G assembly}
                \vspace{2pt}
                \begin{minted}[fontsize=\footnotesize,linenos,xleftmargin=18pt,numbersep=6pt]{asm}
fld     fa3, 0(|\tikzmark{startdep0}|a3|\tikzmark{enddep0}|)
fmul.d  fa3, %[InvLn2N], fa3
fadd.d  fa1, fa3, %[SHIFT]
fsd     fa1, |\tikzmark{startdep1}|0(%[ki])|\tikzmark{enddep1}|
lw      a0, |\tikzmark{startdep2}|0(%[ki])|\tikzmark{enddep2}|
andi    a1, a0, 0x1f
slli    a1, a1, 0x3
add     a1, %[T], a1
lw      a2, 0(a1)
lw      a1, 4(a1)
slli    a0, a0, 0xf
sw      a2, |\tikzmark{startdep3}|0(%[t])|\tikzmark{enddep3}|
add     a0, a0, a1
sw      a0, |\tikzmark{startdep4}|4(%[t])|\tikzmark{enddep4}|
fsub.d  fa2, fa1, %[SHIFT]
fsub.d  fa3, fa3, fa2
fmadd.d fa2, %[C0], fa3, %[C1]
fld     fa0, |\tikzmark{startdep5}|0(%[t])|\tikzmark{enddep5}|
fmadd.d fa4, %[C2], fa3, %[C3]
fmul.d  fa1, fa3, fa3
fmadd.d fa4, fa2, fa1, fa4
fmul.d  fa4, fa4, fa0
fsd     fa4, 0(|\tikzmark{startdep6}|a4|\tikzmark{enddep6}|)
addi    |\tikzmark{startdep7}|a3|\tikzmark{enddep7}|,  a3, 8
addi    |\tikzmark{startdep8}|a4|\tikzmark{enddep8}|,  a4, 8
                \end{minted}
                \vspace{-6pt}
                \subcaption{}
                \vspace{-7pt}
                \label{lst:asm}
                \begin{tikzpicture}[
                    remember picture,
                    overlay,
                    annotation/.style={
                        inner sep=2pt,
                        text width=5cm,
                        rotate=90,
                        align=center,
                        font=\ttfamily\footnotesize
                    },
                    ]

                    \draw[red]
                    ($(pic cs:startdep0)+(-0.15em,0.75em)$) -- 
                    ($(pic cs:enddep0)+(0.15em,0.75em)$) -- 
                    ($(pic cs:enddep0)+(0.15em,-0.25em)$) --  
                    ($(pic cs:startdep0)+(-0.15em,-0.25em)$) -- 
                    cycle;

                    \draw[blue]
                    ($(pic cs:startdep1)+(-0.15em,0.75em)$) -- 
                    ($(pic cs:enddep1)+(0.15em,0.75em)$) -- 
                    ($(pic cs:enddep1)+(0.15em,-0.25em)$) --  
                    ($(pic cs:startdep1)+(-0.15em,-0.25em)$) -- 
                    cycle;

                    \draw[red]
                    ($(pic cs:startdep2)+(-0.15em,0.75em)$) -- 
                    ($(pic cs:enddep2)+(0.15em,0.75em)$) -- 
                    ($(pic cs:enddep2)+(0.15em,-0.25em)$) --  
                    ($(pic cs:startdep2)+(-0.15em,-0.25em)$) -- 
                    cycle;

                    \draw[blue]
                    ($(pic cs:startdep3)+(-0.15em,0.75em)$) -- 
                    ($(pic cs:enddep3)+(0.15em,0.75em)$) -- 
                    ($(pic cs:enddep3)+(0.15em,-0.25em)$) --  
                    ($(pic cs:startdep3)+(-0.15em,-0.25em)$) -- 
                    cycle;

                    \draw[blue]
                    ($(pic cs:startdep4)+(-0.15em,0.75em)$) -- 
                    ($(pic cs:enddep4)+(0.15em,0.75em)$) -- 
                    ($(pic cs:enddep4)+(0.15em,-0.25em)$) --  
                    ($(pic cs:startdep4)+(-0.15em,-0.25em)$) -- 
                    cycle;

                    \draw[red]
                    ($(pic cs:startdep5)+(-0.15em,0.75em)$) -- 
                    ($(pic cs:enddep5)+(0.15em,0.75em)$) -- 
                    ($(pic cs:enddep5)+(0.15em,-0.25em)$) --  
                    ($(pic cs:startdep5)+(-0.15em,-0.25em)$) -- 
                    cycle;

                    \draw[red]
                    ($(pic cs:startdep6)+(-0.15em,0.75em)$) -- 
                    ($(pic cs:enddep6)+(0.15em,0.75em)$) -- 
                    ($(pic cs:enddep6)+(0.15em,-0.25em)$) --  
                    ($(pic cs:startdep6)+(-0.15em,-0.25em)$) -- 
                    cycle;

                    \draw[blue]
                    ($(pic cs:startdep7)+(-0.15em,0.75em)$) -- 
                    ($(pic cs:enddep7)+(0.15em,0.75em)$) -- 
                    ($(pic cs:enddep7)+(0.15em,-0.25em)$) --  
                    ($(pic cs:startdep7)+(-0.15em,-0.25em)$) -- 
                    cycle;

                    \draw[blue]
                    ($(pic cs:startdep8)+(-0.15em,0.75em)$) -- 
                    ($(pic cs:enddep8)+(0.15em,0.75em)$) -- 
                    ($(pic cs:enddep8)+(0.15em,-0.25em)$) --  
                    ($(pic cs:startdep8)+(-0.15em,-0.25em)$) -- 
                    cycle;
                \end{tikzpicture}
            \end{minipage}
        \end{minipage}%
        \vline
        %
        \begin{minipage}[t]{.18\textwidth}
            \vspace{0pt}
            \begin{subfigure}{\textwidth}
                \centering
                \vspace{4pt}
                \textbf{Steps 1 \& 2}\par\medskip
                \includegraphics[width=0.96\textwidth]{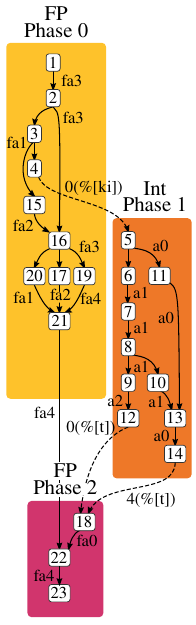}
                \vspace{-1pt}
                \subcaption{}
                \label{fig:dfg}
            \end{subfigure}
        \end{minipage}%
        \vline
        %
        \begin{minipage}[t]{.28\textwidth}
            \vspace{0pt}
            \begin{minipage}[t]{\textwidth}
                \vspace{0pt}
                \centering
                \vspace{4pt}
                \textbf{Step 3}
                \vspace{2pt}
                \begin{minted}[fontsize=\footnotesize,xleftmargin=4pt]{c}
for (int i = 0; i < N; i++) {
  phase0(x[i], &ki, &w);
  phase1(ki, &t);
  phase2(w, t, &y[i]);
}
                \end{minted}
                \vspace{-12pt}
                \subcaption{}
                \vspace{-4pt}
                \label{lst:scheduling}
                \hrulefill
            \end{minipage}

            \begin{minipage}[t]{\textwidth}
                \centering
                \vspace{4pt}
                \textbf{Step 4}
                \vspace{2pt}
                \begin{minted}[fontsize=\footnotesize,xleftmargin=4pt]{c}
for (int j = 0; j < N/B; j++) {
  for (int i = 0; i < B; i++)
    phase0(x[i], &ki[i], &w[i]);
  for (int i = 0; i < B; i++)
    phase1(ki[i], &t[i]);
  for (int i = 0; i < B; i++)
    phase2(w[i], t[i], &y[i]);
}      
                \end{minted}
                \vspace{-12pt}
                \subcaption{}
                \vspace{1em}
                \label{lst:tiling}
            \end{minipage}

            \begin{subfigure}{\textwidth}
                \centering
                \includegraphics[width=0.96\textwidth]{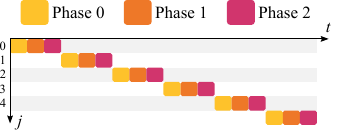}
                \vspace{-8pt}
                \subcaption{}
                \vspace{-4pt}
                \label{fig:tiling}
                \hrulefill
            \end{subfigure}

            \begin{subfigure}{\textwidth}
                \centering
                \vspace{4pt}
                \textbf{Step 5}\par\medskip
                \vspace{-4pt}
                \includegraphics[width=0.96\textwidth]{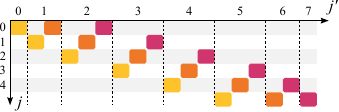}
                \vspace{-4pt}
                \subcaption{}
                \label{fig:pipelining}
            \end{subfigure}
        \end{minipage}%
        \vline
        %
        \begin{minipage}[t]{0.25\textwidth}
            \vspace{0pt}
            \begin{minipage}[t]{\textwidth}
                \centering
                \vspace{4pt}
                \textbf{Step 6}
                \vspace{2pt}
                \begin{minted}[fontsize=\footnotesize,xleftmargin=4pt]{asm}
fld fa3, 0(a3)
      ↓
lw  t0, 0(a3)
lw  t1, 4(a3)
sw  t0, 0(%[tmp])
sw  t1, 4(%[tmp])
fld fa3, 0(%[tmp])
                \end{minted}
                \vspace{-4pt}
                \subcaption{}
                \label{lst:conversion}
            \end{minipage}
            \vspace{4pt}

            \begin{minipage}[t]{\textwidth}
                \centering
                \begin{minted}[fontsize=\footnotesize,xleftmargin=4pt]{python}
for i in range(N):
  addr = i * stride1 + base1
for i in range(N):
  addr = i * stride1 + base2
             ↓
for j in range(2):
  for i in range(N):
    addr = j * (base2-base1) 
         + i * stride1
         + base1
                \end{minted}
                \vspace{-4pt}
                \subcaption{}
                \label{lst:stream-fusion}
                \hrulefill
            \end{minipage}

            \begin{subfigure}{\textwidth}
                \centering
                \vspace{4pt}
                \textbf{Step 7}\par\medskip
                \includegraphics[width=0.96\textwidth]{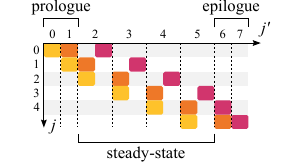}
                \caption{}
                \label{fig:schedule}
            \end{subfigure}
        \end{minipage}
    \end{mdframed}
    \caption{Illustration of the COPIFT methodology applied to a kernel calculating the exponential of a vector.}
\end{figure*}

Akin to other \glspl{isa}, RISC-V defines distinct \glspl{rf} for the RV32I base integer \gls{isa} and the ``D'' standard extension\cite{riscv}. As a result, integer and \gls{fp} instructions operate mostly on independent sets of registers.
This fundamental characteristic allows us to define two threads with independent state, comprising respectively the integer \gls{rf} and RV32I instructions and the \gls{fp} \gls{rf} and ``D'' extension instructions.
Several notable \gls{fp} instructions, however, also require access to the integer \gls{rf}, consuming operands from integer source registers (\gls{fp} load, store, conversion and move instructions) or producing results into integer destination registers (\gls{fp} comparisons).
Such instructions conflict with the independent thread abstraction, precluding pseudo-dual issue execution.

Partly relaxing this constraint, Snitch implements \glspl{ssr} \cite{schuiki2021}, a custom \gls{isa} extension to stream data from memory directly into the \gls{rf}, eliminating explicit load and store instructions.
\glspl{ssr} support any \textit{affine stream} or memory access pattern which can be described as an affine function of the induction variables of a four-dimensional loop nest.
Extending the \gls{ssr} mechanism, \glspl{issr} \cite{scheffler2023}, further provide support for \textit{indirect streams} or arbitrary memory access patterns, provided a list of relative addresses encoding the pattern.
\glspl{ssr} enable memory accesses without introducing dependencies on the integer \gls{rf}, solving only a particular instance of the problem.
Conversely, our methodology tackles any form of interdependency between integer and \gls{fp} threads.

Consider the code in \Cref{lst:c}, computing the exponential of an array of single-precision \gls{fp} numbers.
Compiling this code for the RV32G architecture results in the mix of integer and \gls{fp} instructions shown in \Cref{lst:asm}.
Dependencies between these instructions are marked with red and blue boxes in the consuming and producing instructions, respectively.
These can be classified into:
\begin{enumerate}[label=\textbf{Type \arabic*},ref=\arabic*,leftmargin=4em,series=types]
    \item \label[type]{type:dynamic-dep} Dynamic memory dependencies, from \gls{fp} load-stores with dynamically-computed addresses.
    \item \label[type]{type:static-dep} Static memory dependencies, from \gls{fp} load-stores with statically-determined addresses.
\end{enumerate}
In the most general case, a third type may be present:
\begin{enumerate}[label=\textbf{Type \arabic*},ref=\arabic*,leftmargin=4em,resume=types]
    \item \label[type]{type:register-dep} Register dependencies, from \gls{fp} conversion, move and comparison instructions.
\end{enumerate}

Due to these dependencies, we cannot trivially partition integer and \gls{fp} instructions for pseudo dual-issue execution.
The COPIFT method addresses the problem by eliminating or increasing the distance between dependent instructions.
It involves the following steps:
\begin{enumerate}[label=\textbf{Step \arabic*},ref=\arabic*,leftmargin=4em]
    \item Construct a \gls{dfg} from the RISC-V assembly and identify all dependencies between integer and \gls{fp} instructions.
    \item \label[step]{step:partitioning} Partition the \gls{dfg} on the identified edges, such that there exists an acyclic precedence-relation among the subgraphs, and the number of edges between them is minimized. 
    \item \label[step]{step:scheduling} Reorder the instructions according to the subgraph ordering established in \Cref{step:partitioning}.
    \item \label[step]{step:tiling} Apply loop tiling to block the data, and loop fission to reschedule the subgraph computations on each block.
    \item \label[step]{step:pipelining} Apply software pipelining and multiple buffering to overlap subgraph computations.
    \item \label[step]{step:ssr} Map all \gls{fp} load-store instructions to \glspl{ssr}.
    \item \label[step]{step:frep} Map all \gls{fp} loops to \gls{frep} loops, and reorder them to preceed the integer loop.
\end{enumerate}
To clarify each step, we walk through how to apply COPIFT to the exponential kernel in \Cref{lst:c}.

\Cref{fig:dfg} shows the \gls{dfg} of the basic block in \Cref{lst:asm}. For brevity, we omitted instructions 24 and 25 and associated dependencies, as we will see that these can be simplified in \Cref{step:ssr}.
Following \Cref{step:partitioning}, we cut edges \edge{4}{5}, \edge{12}{18} and \edge{14}{18}.
This alone results in two subgraphs for which a precedence-relation cannot be established.
We therefore further partition the graph, cutting edge \edge{21}{22}.
We obtain three subgraphs, each defining a phase of the computation with clear ordering requirements w.r.t. the others: \gls{fp} Phase 0 must precede integer Phase 1, which in turn must precede \gls{fp} Phase 2.
This allows us to reorder the instructions in consecutive groups of integer-only or \gls{fp}-only instructions, while respecting the dependencies within each loop iteration (\Cref{step:scheduling}).
The resulting code would resemble \Cref{lst:scheduling}, where each function would be replaced by the respective inline assembly code.

In \Cref{step:tiling}, we then apply loop tiling \cite{hammami2017} to obtain the code in \Cref{lst:tiling}.
Each phase is now applied to an entire block of inputs, before passing on a block of intermediate results to the next phase.
Implicitly, this step requires allocating block-sized buffers for each variable associated to an edge between subgraphs.
Additionally, all inter-phase communication originally through the \gls{rf}, must be preventively spilled to memory.
To reduce spill communication overheads and memory requirements, it is important to minimize the number of edges between subgraphs in \Cref{step:partitioning}.

\Cref{step:pipelining} prepares for the parallel execution of the \gls{fp} and integer phases, by applying software pipelining \cite{douillet2007} to increase the distance between dependent phases.
\Cref{fig:tiling,fig:pipelining} illustrate the original and modified block schedules, respectively.
Within each $j'$ iteration in the new schedule, every phase operates on different blocks of data $j$, allowing them to be parallelized in \Cref{step:frep}.
Implicitly, this requires replicating the buffers, as e.g. Phase 0 produces different blocks of $w$ before the first is consumed by Phase 2.
The exact number of replicas for each buffer equals the distance between the subgraphs connected by the respective edge in the total ordering established in \Cref{step:partitioning}, plus one.
In our example, the $w$ buffer, associated to the edge between Phase 0 and 2, must be replicated three times.

\begin{table*}[t!]
    \begin{tabular*}{\textwidth}{@{\extracolsep{\fill}} l c c c c c c c c c c c c c @{}}
        \toprule
        \multirow{3}{*}{\makecell{Kernel}} & \multicolumn{3}{c}{Baseline} & \multicolumn{2}{c}{\Cref{step:tiling}} & \multicolumn{2}{c}{\Crefrange{step:pipelining}{step:ssr}} & \multirow{3}{*}{\makecell{Max \\ Block}} & \multicolumn{2}{c}{COPIFT} & \multirow{3}{*}{\makecell{$I'$}} & \multirow{3}{*}{\makecell{$S''$}} & \multirow{3}{*}{\makecell{$\blacktriangledown S'$}} \\
        \cmidrule(lr){2-4}
        \cmidrule(lr){5-6}
        \cmidrule(lr){7-8}
        \cmidrule(lr){10-11}
        & \makecell{\#Int} & \makecell{\#FP} & \makecell{TI} & \makecell{Int Ld/St} & \makecell{\#Buff.} & \makecell{FP Ld/St} & \makecell{\#Buff.} & & \makecell{\#Int} & \makecell{\#FP} & & & \\
        \midrule
        \texttt{expf}                                     & 43  & 52 & 0.83 & 0  & 5 & -4 & 13 & 157 & 43  & 36 & 1.84 & 1.83 & 2.21 \\
        \texttt{logf}*\textsuperscript{\ddag}             & 39  & 52 & 0.75 & +4 & 6 & -4 & 12 & 273 & 57  & 36 & 1.63 & 1.75 & 1.6  \\
        \texttt{poly\_lcg}*\textsuperscript{\dag}         & 44  & 80 & 0.55 & +3 & 3 & 0  & 6  & 341 & 72  & 80 & 1.9  & 1.55 & 1.55 \\
        \texttt{pi\_lcg}*\textsuperscript{\dag}           & 44  & 56 & 0.79 & +3 & 3 & 0  & 6  & 341 & 72  & 56 & 1.78 & 1.79 & 1.39 \\
        \texttt{poly\_xoshiro128p}*\textsuperscript{\dag} & 172 & 80 & 0.47 & +3 & 3 & 0  & 6  & 341 & 200 & 80 & 1.4  & 1.47 & 1.26 \\
        \texttt{pi\_xoshiro128p}*\textsuperscript{\dag}   & 172 & 56 & 0.33 & +3 & 3 & 0  & 6  & 341 & 200 & 56 & 1.28 & 1.33 & 1.14 \\
        \bottomrule
        \addlinespace[2pt]
        \multicolumn{13}{l}{Require support for *\texttt{fcvt.d.w} and \textsuperscript{\dag}\texttt{flt.d}; \textsuperscript{\ddag}use \glspl{issr} to map \Cref{type:dynamic-dep} dependencies.}
    \end{tabular*}
    \caption{Characteristics of the evaluated kernels, ordered by expected speedup $S'$.}
    \label{tab:kernels}
\end{table*}

\Crefrange{step:ssr}{step:frep} enable the integer and \gls{fp} phases within each block iteration to be executed in parallel.
In \Cref{step:tiling}, we spilled all inter-phase communication through the \glspl{rf} to memory, introducing additional load-store instructions.
In \Cref{step:ssr}, we eliminate all \gls{fp} load-stores by mapping the respective memory accesses to \glspl{ssr}.
The accesses performed by the two \gls{fp} phases in our example can be described as a combination of 6 one-dimensional streams: reads from blocks of $x$, $w$ and $t$, and writes to blocks of $w$, $ki$ and $y$.
All streams originate from tiling in \Cref{step:tiling}, and can thus be naturally represented as regular accesses into contiguous arrays.
Special treatment is required to convert streams of addresses originating from \Cref{type:dynamic-dep} dependencies, into regular streams of data for the \gls{fp} thread.
This can be done in software by converting \Cref{type:dynamic-dep} to \Cref{type:static-dep} dependencies as shown in \Cref{lst:conversion}.
This involves prefetching the data in the integer thread and storing it in a regular layout for the \glspl{ssr}.
Alternatively, \Cref{type:dynamic-dep} dependencies can be directly mapped to \glspl{issr}, performing the indirection on the regular stream of addresses in hardware.
Finally, we note that in cases where the \Cref{type:dynamic-dep} address stream can be statically generated by an affine function of up to four variables, the entire address calculation can be eliminated and performed by an \gls{ssr}, which is the case for the $x$ and $y$ streams in our example.
The associated dependency and all artifacts produced by our methodology can be optimized away, which allowed us to ignore instructions 24 and 25 from the start.

To map the 6 streams to the 3 \glspl{ssr} available in every Snitch core, we apply a \textit{stream fusion} technique, which involves mapping multiple lower-dimensional affine streams to a single higher-dimensional affine stream.
\Cref{lst:stream-fusion} illustrates the process of merging two one-dimensional affine streams.
Using this technique, we merge the $x$ and $t$ read streams and the $w$, $ki$ and $y$ write streams.

Finally, \gls{fp} block computations are mapped to \gls{frep} loops (\Cref{step:frep}).
Since instructions for the first iteration of an \gls{frep} loop are issued by the integer core, the \gls{frep} loops must preceed the integer loop in program order, such that remaining iterations overlap with the integer thread.
Similarly, for the integer thread to overlap with both \gls{fp} Phase 0 and 2, these should be fused to a single \gls{frep} loop in block iterations where both phases are performed.
The resulting schedule is shown in \Cref{fig:schedule}.

The steps in this methodology can be followed by developers to obtain an optimized mixed C and assembly representation.

\subsection{COPIFT ISA Extensions}
\label{sec:extensions}

\gls{fp} conversion and comparison instructions are critical for mixed integer and \gls{fp} workloads.
To support these, we can alter their semantics under \gls{frep} operation, to operate entirely on the \gls{fp} \gls{rf}.
Communication between the integer and \gls{fp} \glspl{rf} is spilled to memory, through the use of additional load-store instructions, decoupling it from the computation, so the target instructions can find all operands directly in the \gls{fp} \gls{rf}.
This technique can be combined with \glspl{ssr}, to implicitly encode the \gls{fp} load-store operations.

Albeit not strictly necessary, we chose to develop custom instructions to carry the altered semantics, for each of the relevant ``D'' extension instructions, i.e. \texttt{fcvt.w[u].d}, \texttt{fcvt.d.w[u]}, \texttt{feq.d}, \texttt{flt.d}, \texttt{fle.d} and \texttt{fclass.d}.
We copy the original encodings, allocating the new instructions in the \texttt{custom-1} opcode space\cite{riscv}.

\section{Results}
\label{sec:results}

We implement a Snitch cluster hosting one compute core in GlobalFoundries' 12LP+ FinFET technology using Fusion Compiler, with a target clock frequency of 1\,GHz.
Our extensions introduce negligible area and timing overheads, within the margin of synthesis process variability.

All experiments are conducted at the target frequency in cycle-accurate RTL simulations using QuestaSim 2023.4.
Switching activities are extracted from post-layout simulations, and used for power estimation in PrimeTime, assuming typical operating conditions of 25\,°C and 0.8\,V supply voltage.

\subsection{Performance Evaluation}

To evaluate the benefits of the COPIFT method and \gls{isa} extensions, we implement a set of mixed integer and \gls{fp} workloads with diverse characteristics, extracted from two different domains:
\begin{itemize}
    \item Hit and miss Monte Carlo integration methods \cite{toral2014}, evaluating two different integration problems and pseudo-random number generators. Integer instructions are typically employed in pseudo-random number generation, while the integration is performed in double-precision \gls{fp} arithmetic.
    \item Transcendental function evaluation on a vector of single-precision \gls{fp} numbers. Reference implementations are taken from the GNU C Library\cite{glibc} v2.40, and involve integer instructions for bit-manipulation and double-precision \gls{fp} arithmetic to evaluate approximating polynomials.
\end{itemize}
The latter in particular, applied to the exponential function, is the main component of softmax operations, which consume a considerable fraction of cycles in modern \glspl{llm} \cite{stevens2021}.

We implement Snitch-optimized RV32G-compliant baseline codes and COPIFT-accelerated variants for each kernel, compiled using a custom toolchain for Snitch based on LLVM 18.1.4 with \lstinline{-O3} optimization.

\Cref{tab:kernels} lists all of the evaluated kernels, with their unique characteristics.
As the table shows, the number of integer and \gls{fp} instructions can vary as a result of applying the COPIFT method, as e.g. integer load-stores are introduced in \Cref{step:tiling} and \gls{fp} load-stores are eliminated in \Cref{step:ssr}.
These easily measurable kernel characteristics can be used to estimate the speedup potential of the COPIFT method.
In fact, assuming similar IPCs in the integer and \gls{fp} threads, the speedup $S$ can be approximated as a function of the number of integer and \gls{fp} instructions:
\begin{equation}
    \begin{split}
        S \coloneqq \frac{t^{\text{base}}_{\text{int}} + t^{\text{base}}_{\text{fp}}}{\max(t^{\text{copift}}_{\text{int}}, t^{\text{copift}}_{\text{fp}})}
          \approx \frac{n^{\text{base}}_{\text{int}} + n^{\text{base}}_{\text{fp}}}{\max(n^{\text{copift}}_{\text{int}}, n^{\text{copift}}_{\text{fp}})} = S'
    \end{split}
    \label{eq:speedup}
\end{equation}
Similarly, we can approximate the \gls{ipc} improvement $I$:
\begin{equation}
    \begin{split}
        I \coloneqq S * \frac{n^{\text{copift}}_{\text{int}} + n^{\text{copift}}_{\text{fp}}}{n^{\text{base}}_{\text{int}} + n^{\text{base}}_{\text{fp}}}
          \approx \frac{n^{\text{copift}}_{\text{int}} + n^{\text{copift}}_{\text{fp}}}{\max(n^{\text{copift}}_{\text{int}}, n^{\text{copift}}_{\text{fp}})} = I'
    \end{split}
\end{equation}
If the number of instructions does not differ significantly between the two code variants, we can further approximate the speedup as a function of the number of instructions in the original implementation alone:
\begin{equation}
    \begin{split}
        S' \approx \frac{n^{\text{base}}_{\text{int}} + n^{\text{base}}_{\text{fp}}}{\max(n^{\text{base}}_{\text{int}}, n^{\text{base}}_{\text{fp}})}
        &\stackrel{\footnotemark}{=} 1 + \frac{\min(n^{\text{base}}_{\text{int}}, n^{\text{base}}_{\text{fp}})}{\max(n^{\text{base}}_{\text{int}}, n^{\text{base}}_{\text{fp}})} = \\
        &= 1 + \mathrm{TI} = S'' = I''
    \end{split}
    \label{eq:imbalance}
\end{equation}
\footnotetext{Using equality $a + b = \max(a, b) + \min(a, b)$}%
Intuitively, this shows how the speedup approximately depends on the \gls{ti}, as defined above.

\begin{figure}[t]
    \begin{subfigure}{\columnwidth}
        \centering
        \includegraphics[width=\textwidth]{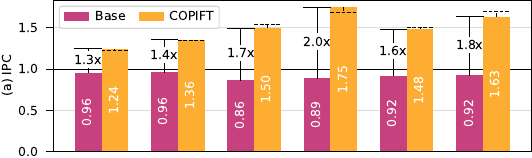}
        \phantomcaption
        \label{fig:ipc}
    \end{subfigure}
    \vspace{-1em}

    \begin{subfigure}{\columnwidth}
        \centering
        \includegraphics[width=\textwidth]{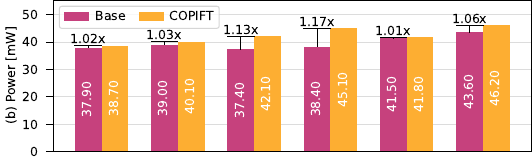}
        \phantomcaption
        \label{fig:power}
    \end{subfigure}
    \vspace{-1em}
    \vspace{-2pt}

    \begin{subfigure}{\columnwidth}
        \centering
        \includegraphics[width=\textwidth]{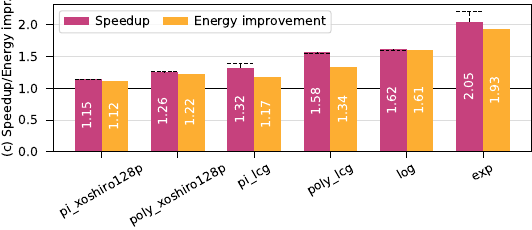}
        \phantomcaption
        \label{fig:speedup}
    \end{subfigure}
    \vspace{-1em}

    \caption{Comparison of various metrics between COPIFT and baseline codes. Dashed lines indicate expected values: (a) the \gls{ipc} derived from $I'$ and (c) the speedup $S'$.}
\end{figure}

\Cref{fig:ipc} compares the \gls{ipc} of a steady-state iteration in the two implementations, for all kernels ordered by increasing expected speedup $S'$.
We observe a significant geomean \gls{ipc} improvement of \ResultGeomeanIPCIncrease$\times$, and a peak \gls{ipc} of \ResultPeakIPC, as a demonstration of dual-issue performance.
As hypothesized, the \gls{ipc} improvement correlates with $I'$.
The deviations observed in the \texttt{poly\_lcg} and \texttt{pi\_lcg} kernels are due to the hypothesis made in \Cref{eq:speedup} being invalid.
In fact, both benchmarks present stalls in the \gls{prn} generation with the \gls{lcg}, which are due to structural hazards on the register file's writeback port, and could not be eliminated by unrolling.
These stalls effectively balance out the execution times of the integer and \gls{fp} threads in the \texttt{poly\_lcg} kernel, leading to a near-ideal \gls{ipc} improvement of \ResultPolyLCGIPCIncrease$\times$, while unbalancing the \texttt{pi\_lcg} kernel.
Finally, the \texttt{exp} kernel also shows a small deviation.
This is due to the instruction overhead required to program the \glspl{ssr} and switch buffers in every block iteration, which is not amortized as effectively on smaller blocks, as this kernel features.

\begin{figure}[t]
    \centering
    \includegraphics[width=\columnwidth]{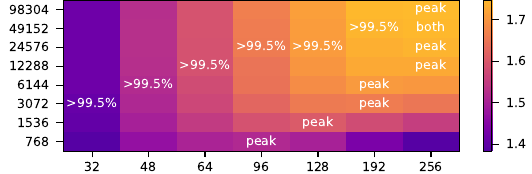}
    \caption{\gls{ipc} of the \texttt{poly\_lcg} kernel for various problem and block sizes.}
    \label{fig:heatmap}
\end{figure}

To understand the impact of block size and problem size on performance, in \Cref{fig:heatmap} we plot the \gls{ipc} for various configurations of the \texttt{poly\_lcg} kernel.
As expected, \gls{ipc} increases with problem size, as fixed prologue and epilogue overheads are amortized over a larger number of block iterations.
Smaller blocks result in smaller prologue and epilogue overheads and a larger number of block iterations, for equal problem sizes, and can thus amortize these overheads more effectively.
This trend is displayed in the ``>99.5\%'' annotations, marking the smallest problem size which achieves >99.5\% of the maximum \gls{ipc} attainable with each block size.
On the other hand, smaller blocks are less able to amortize overheads within the block, e.g. from \gls{ssr} programming and switching buffers.
As such, smaller block sizes feature lower maximum \glspl{ipc}, when all other overheads are fully amortized.
In this tradeoff, there exists an optimal block size per problem size, which we annotate with the ``peak'' labels.
As the problem size increases, the peak shifts towards larger block sizes, up to the maximum size we can fit in L1 memory, reported in \Cref{tab:kernels}.
Finally, we note that, as we tend to amortize all overheads, the \gls{ipc} converges to the steady-state \gls{ipc} presented in \Cref{fig:ipc}.
In the following discussions we will thus restrict our analysis to steady-state measurements.

\Cref{fig:speedup} shows how the \gls{ipc} increase translates to improved kernel performance.
Again, we observe a significant geomean speedup of \ResultGeomeanSpeedup$\times$, with a peak \ResultPeakSpeedup$\times$ speedup on the \texttt{exp} kernel.
The deviations from the expected speedup $S'$ are a consequence of the same overheads affecting the \gls{ipc}.
Finally, we note that speedups greater than two are possible, as a result of additional optimizations, such as load-store elision with the \glspl{ssr}, on top of dual-issue execution.

\subsection{Energy Evaluation}


\Cref{fig:power} compares the power consumption of the two implementations.
The Monte Carlo benchmarks exhibit a lower base power consumption than their counterparts, due to the inactivity of the \gls{dma} engine and reduced L1 memory accesses.
The increase in power is consistently correlated with the \gls{ipc} increase, with exception of the \texttt{log} and \texttt{exp} kernels.
In fact, after separating the \gls{fp} instructions, the integer loop bodies of these two kernels comprise less than 64 instructions, which entirely fit in Snitch's L0 I\$.
As a result, the I\$ consumes significantly less power in the COPIFT implementations, which do not experience thrashing at the L0 level.

Ultimately, as the power consumption is dominated by constant components such as the clock network activity, we observe a maximum power increase of only \ResultMaxPowerIncrease$\times$, and a geomean power increase of \ResultGeomeanPowerIncrease$\times$ over all kernels.
As a result, COPIFT is still largely beneficial in overall energy terms, as shown in \Cref{fig:speedup}.
The reduced execution time, resulting from dual-issue execution, outweighs the increase in power consumption, resulting in geomean energy savings of \ResultGeomeanEnergySaving$\times$ on the measured benchmarks, and a peak \ResultPeakEnergySaving$\times$ saving on the \texttt{exp} kernel.

\section{Related Work}


To issue multiple instructions in parallel while preserving program correctness, multiple-issue processors must be able to identify dependencies within an instruction stream.
Multiple-issue capabilities are often built into out-of-order processors, leveraging the facilities which enable out-of-order execution, such as reservation stations, register renaming and reorder buffers, which lead to high area and power overheads.
On the other hand, in-order multiple-issue processors can be implemented in two flavours: statically-scheduled superscalar processors and VLIW processors \cite{hennessy2011}.
In-order superscalar processors integrate logic to identify hazards between instructions in the issue packet and those in execution, while VLIW processors offload part of the job to the compiler, in charge of packing independent instructions into bundles of a predetermined size, at the cost of increased code size.

Few recent works in the literature look into the area- and energy-efficient design of dual-issue in-order RISC processors.
All of these target dual-issue execution of arbitrary instruction pairs, with minor restrictions; as a consequence, they require a minimum of four read and two write (4R2W) ports into the \gls{rf}.
Lozano et al. \cite{lozano2015} design a novel latency-optimized 4R2W-port SRAM-based \gls{rf} implementation for soft FPGA cores, yet their implementation incurs a 4$\times$ cost in physical RAM blocks compared to the single-issue baseline, which also accounts for the majority of the +18\% power increase.
Similarly, Kra et al. \cite{kra2024} implement a dual-issue in-order RISC-V core based on CV32E40P, duplicating instruction decode, issue and execution stages and integrating a 5R3W-port \gls{rf} with cross-forwarding support from all write to all read ports, for a total 60\% area increase.
Patsidis et al. \cite{patsidis2018} claim to further push the area and energy efficiency limits of dual-issue RISC-V cores, by restricting dual-issue to 16-bit compressed instructions, fitting within a single 32-bit instruction fetch port.
They extend the \gls{rf} with extra read ports, and a partitioned design to harvest the energy benefits from locality of reference to eight commonly-addressed registers by compressed instructions.

Wygrzvwalski et al. \cite{wygrzvwalski2024} present a pseudo dual-issue mechanism for small loop bodies, which resembles the Snitch implementation.
A separate instruction buffer is allocated as a source of instructions for the second issue slot.
Differently from Snitch, they rely on a banked \gls{rf} design to prevent structural hazards on the \gls{rf} ports, restricting parallel execution to pairs of instructions accessing different register banks.
Conversely, Snitch targets pseudo dual-issue execution of integer and \gls{fp} instructions, leveraging the separate \glspl{rf} available by \gls{isa} design.
Unfortunately, Wygrzvwalski's design is targeted to a specific application, limiting their investigation to a rather simple code featuring two independent threads operating in lockstep, without diving into the intricacies required to support more complex general-purpose scenarios.

In contrast to these works, we presented a novel methodology enabling dual-issue execution of mixed integer and \gls{fp} instruction sequences in an area- and energy-efficient processor, with virtually no area overhead due to our extensions.
This is possible by leveraging Snitch's \gls{ssr} and \gls{frep} extensions, which extend beyond the scope of dual-issue execution, enabling the core's high \gls{fpu} utilization.

\section{Conclusion}

In this work, we presented COPIFT, a generic methodology to enable sustained dual-issue execution of mixed integer and floating-point instruction sequences in Snitch, a state-of-the-art open-source RISC-V processor.
We further developed a set of \gls{isa} extensions to broaden the range of applications which can leverage our methodology.
To evaluate the benefits of COPIFT-enabled dual-issue execution, we implemented optimized RV32G baseline codes for a variety of mixed integer and floating-point workloads on Snitch, and compared these to COPIFT-optimized variants leveraging our \gls{isa} extensions.
We measured speedups of \ResultGeomeanSpeedup$\times$ on average and a peak \gls{ipc} of \ResultPeakIPC{}, proving COPIFT's effectiveness in enabling sustained dual-issue execution.
We further compared the energy-efficiency of baseline and COPIFT-accelerated codes.
Despite a small \ResultGeomeanPowerIncrease$\times$ average increase in power consumption, COPIFT proves beneficial in overall energy terms, resulting in \ResultGeomeanEnergySaving $\times$ average energy savings on the measured benchmarks, and a peak \ResultPeakEnergySaving $\times$ saving on the \texttt{exp} kernel, proving that effective dual-issue execution is possible on area- and energy-efficient in-order cores. 

\section*{Acknowledgment}

\ifdefined\blindreview
Hidden for double blind review.
\else
This work has been supported in part by ‘The European Pilot’ project under grant agreement No 101034126 that receives funding from EuroHPC-JU as part of the EU Horizon 2020 research and innovation programme.

We would like to thank Lannan Jiang, Fabio Cappellini and Hakim Filali for their early experiments with Monte Carlo methods on Snitch. 
\fi

\bibliography{paper}
\bibliographystyle{IEEEtran}

\end{document}